\documentstyle[epsf]{mn}

\title[C$\rm ^{18}$O observations and modelling of L1689B]
{The initial conditions of isolated star formation: IV -- C$^{18}$O
observations and modelling of the pre-stellar core L1689B}
\author[N. E. Jessop and D. Ward-Thompson]
       {N. E. Jessop $^1$ and D. Ward-Thompson$^2$ \\
       $^1$ Joint Astronomy Center, 660, N. A`Ohoku Place, Hilo, Hawaii, USA\\
       $^2$ Department of Physics and Astronomy, Cardiff University,
            P.O.Box 913, Cardiff, UK}

\date{Accepted 2000 October 25. Received 2000 October 24;
in original form 2000 April 19.}

\pagerange{\pageref{firstpage}--\pageref{lastpage}}

\begin{document}

\maketitle

\label{firstpage}

\begin{abstract}
We present C$\rm ^{18}$O observations of the pre-stellar core L1689B, in the
(J=3$ \rm \rightarrow$2) and (J=2$ \rm \rightarrow$1) rotational transitions,
taken at the James Clerk Maxwell Telescope in Hawaii. We use a 
$\lambda$-iteration 
radiative transfer code to model the data. We adopt a similar
form of radial density profile to that which we have found in all pre-stellar
cores, with a `flat' inner profile, steepening towards the edge, but we make
the gradient of the `flat' region a free parameter. We find that the core
is close to virial equilibrium, but there is tentative 
evidence for core contraction. We allow the temperature 
to vary with a power-law form and find we can consistently fit all of
the CO data with an inverse temperature gradient that is warmer at the
edge than the centre. However, when we combine the CO data with the
previously published millimetre data we fail to find a simultaneous fit
to both data-sets
without additionally allowing the CO abundance to decrease towards
the centre. This effect has been observed qualitatively many times before,
as the CO freezes out onto the dust grains at high densities, but we
quantify the effect.
Hence we show that the combination of mm/submm continuum and
spectral line data is a very powerful method of constraining the physical
parameters of cores on the verge of forming stars.
\end{abstract}


\section{Introduction}

Star formation occurs in dense molecular cloud cores, and many surveys
of such regions have previously been carried out, including the pioneering
work of Myers and co-workers (e.g. Benson \& Myers 1989 and references 
therein). They separated these cloud cores into those that had already 
formed stars and thus contain embedded Young Stellar Objects (YSOs), and
those that had not -- the so-called `starless cores' (Beichman et al 1986).
The starless cores are prime candidates to study observationally the sites 
of potential future star formation, as they are believed to represent the 
initial conditions for protostellar collapse. We have been observing starless
cores for a number of years to try to constrain theoretical models of
protostellar collapse.

Ward-Thompson et al. (1994 -- hereafter Paper I) 
showed that many starless cores contain dense
central condensations which they named `pre-protostellar cores' (or more
recently `pre-stellar cores' for brevity).
Detailed observational studies of pre-stellar cores offer the opportunity
to ascertain the density and temperature distribution
within the core, as well as the kinematics and details of the chemistry,
including dust-molecule interactions.
All of these factors are thought to play
important roles in governing the way in which a body of gas collapses to form
a protostar. 

In Paper I we found that the variation of density ($\rho$) with radius 
($r$) in pre-stellar cores
is very different from the singular isothermal sphere
($\rho \propto r^{-2}$ everywhere) originally
suggested by Shu (1977) as the initial conditions for star formation.
Instead the cores appear to have a much flatter density
profile in the inner region ($\rho \propto r^{-1}$), 
steepening towards their edges ($\rho \propto r^{-2}$).
This was subsequently confirmed for the 
pre-stellar core L1689B by Andr\'e, Ward-Thompson
\& Motte (1996 -- hereafter Paper II), and for other cores
by Ward-Thompson, Motte \& Andr\'e (1999 -- hereafter Paper III).
Most recently, an ISOCAM study by Bacmann et al. (2000) has shown that
some cores have very steep edges indeed ($\rho \propto r^{-5}$).
In a study of one prestellar core (L1544), Tafalla et al. (1998)
discovered significant large-scale motions of gas, possibly indicating
contraction of the core.

\begin{figure*}
\setlength{\unitlength}{1mm}
\begin{picture}(80,130)
\includegraphics{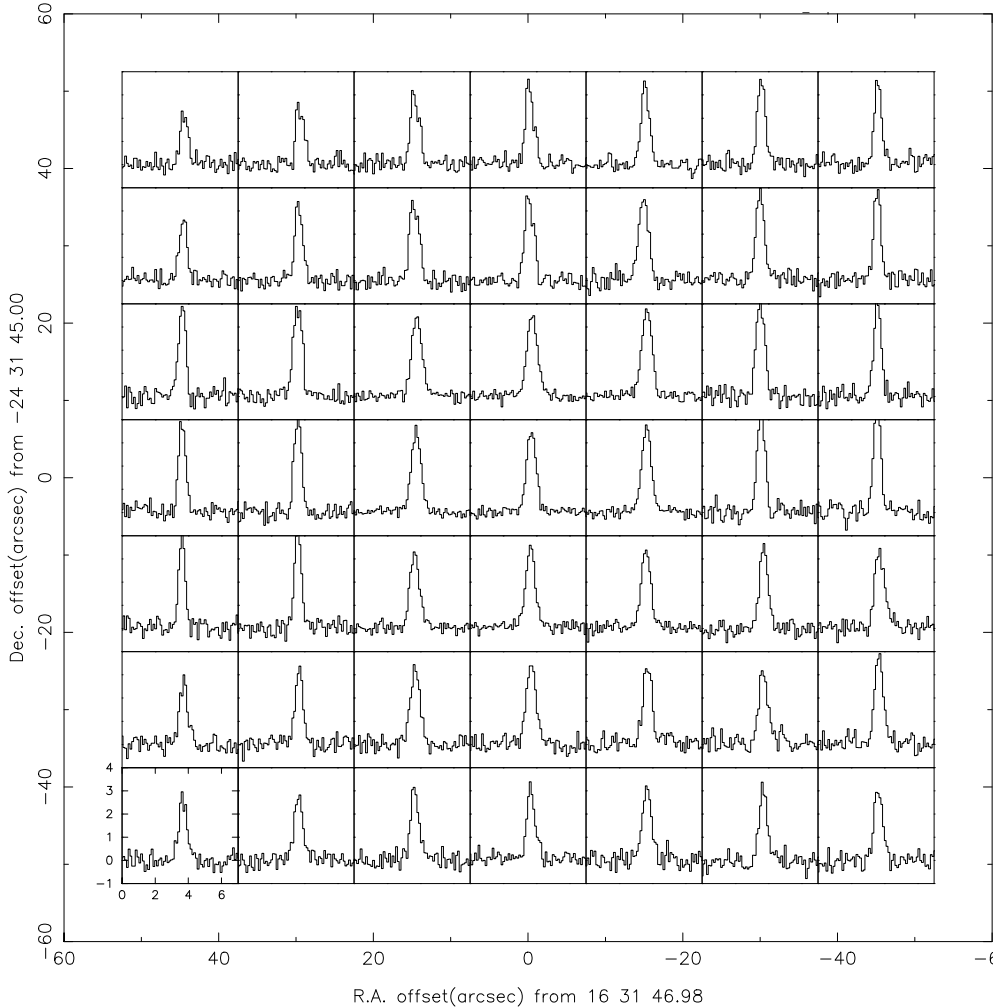}
\end{picture}
\caption{L1689B as seen in the C$^{18}$O (J=2$\rightarrow$1)
rotational transition. T$_{mb}/K$ is plotted against $v/kms^{-1}$ at
each position. Axes are labelled in 1950 co-ordinates. See text for
discussion.}
\end{figure*}

In this paper we present a C$ \rm ^{18}$O study of one pre-stellar
core, L1689B, which has been previously studied at 1.3mm 
(Paper II), and by using a
detailed model of the radiative transfer, we investigate the
physical parameters of this core.
The remainder of the paper is laid out as follows:
Section 2 presents the
C$ \rm ^{18}$O observations of L1689B.
Section 3 presents a radiative transfer model of L1689B.
In Section 4 the variation of density and temperature of the gas
is explored, and we introduce the 1.3mm data to show how
that leads to further constraints, including the variation
of the abundance of CO relative to the dust. Section 5 summarises the
main conclusions.

\section{C$^{18}$O data}

\subsection{Observations}

The observations were carried out at the James Clerk Maxwell Telescope
(JCMT)\footnote{JCMT 
is operated by the Joint Astronomy Center, Hawaii, on 
behalf of the UK PPARC, the Netherlands NWO, and the Canadian NRC.}, 
located on Mauna Kea, Hawaii, on 1995 July 18th,
22nd and 25th
17:30--01:30 {\small HST} ({\small UT} = 03:30--11:30), on 1996 August 31st
and September 1st 17:30--01:30 {\small HST} ({\small UT} = 03:30--11:30).

\begin{figure*}
\setlength{\unitlength}{1mm}
\begin{picture}(80,100)
\includegraphics{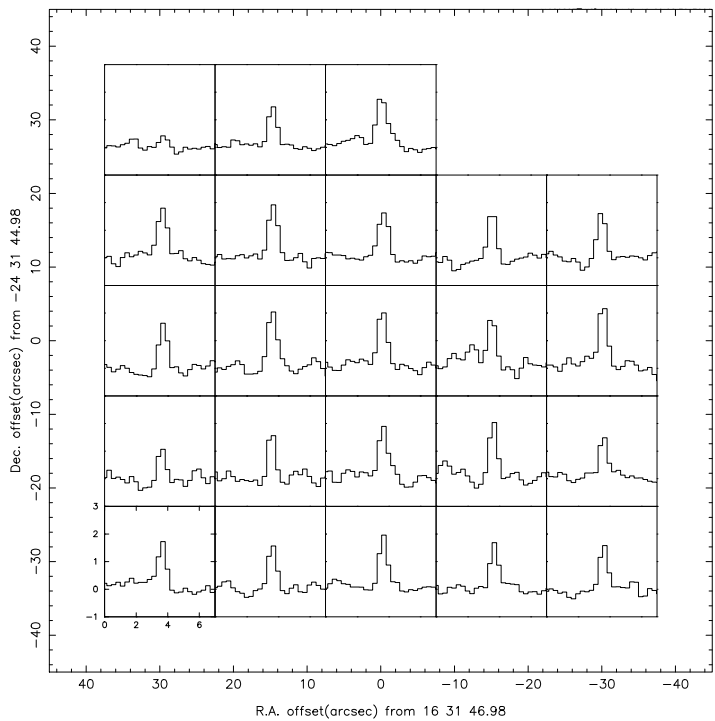}
\end{picture}
\caption{L1689B as seen in the C$^{18}$O (J=3$\rightarrow$2)
rotational transition. T$_{mb}/K$ is plotted against $v/kms^{-1}$ at
each position. Axes are labelled in 1950 co-ordinates. See text for
discussion.}
\end{figure*}

The C$ \rm ^{18}$O (J=3$ \rm \rightarrow$2) transition, with
a rest frequency of
329.33 GHz, and the C$ \rm ^{18}$O (J=2$ \rm \rightarrow$1) transition,
with a rest frequency of 219.56 GHz,
were observed using the common user heterodyne receivers RxB3i 
(Cunningham et al. 1992)
and RxA2 (Davies et al. 1992).
The JCMT half-power beam-width
(HPBW) is 19 arcsec at 220~GHz and 14 arcsec at 329~GHz.
Double-sideband system temperatures
were 2000--12000~K for receiver B3i and 480K for 
receiver A2 observations. 
The backend used was a digital auto-correlation
spectrometer (DAS), with a resolution of 378kHz and 95 kHz per channel,
for the J=(3$ \rm \rightarrow$2) and J=(2$ \rm \rightarrow$1)
transitions respectively, corresponding to
0.34 kms$ \rm ^{-1}$ at J=(3$ \rm \rightarrow$2) and
0.13 kms$ \rm ^{-1}$ at J=(2$ \rm \rightarrow$1).
Data reduction was carried out using the {\small SPECX} package
(Padman 1990). The weather during the observations was in general good for
millimetre (RxA2) observations and short periods proved adequate for
observations of the submillimetre 
J=(3$ \rm \rightarrow$2) transition.

Regular pointing checks on bright sources, observed with the continuum 
backend system, were carried out throughout the observing run
to check the performance of the telescope. 
We found that at worst the pointing was accurate to within
$\sim$3 arcsec. Once 
the receivers had been tuned to the correct frequency, a 
bright standard source was observed to check the chopper wheel calibration.

\subsection{Data Reduction}

The data obtained were in the form of a 
series of spectra sampled at different positions that could be built into
a data cube. The 
basic data format is a spectrum whose intensity is given in terms of the
antenna temperature $ \rm T_a^*$. The spectra 
were calibrated regularly using the usual chopper wheel
method (Kutner \& Ulich 1981) in addition to the standard observations
described above.

\begin{figure*}
\setlength{\unitlength}{1mm}
\begin{picture}(80,90)
\includegraphics{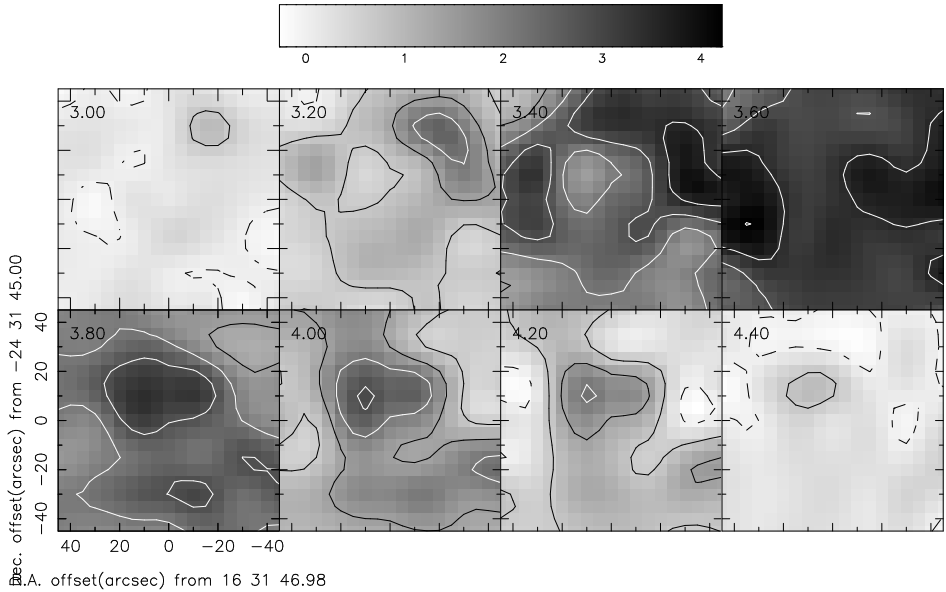}
\end{picture}
\caption{A channel map of the L1689B in the C$ \rm ^{18}$O (J=2$ \rm 
\rightarrow$1) transition. Each map in the sequence shows the integrated
intensity of a 0.2km/s range in velocity. The line centre is at 3.7 km/s.
The red-shifted maps peak on centre. However the blue shifted maps peak
to the east and west, due to a slight shifting of the position of the line
centre at these positions. No evidence for rotation exists. The scale bar
shows the gray scale in units of K.}
\end{figure*}

Since the beam of the telescope is not expected to be
perfectly coupled to the source, a 
correction factor $\eta$ has to be applied to $ \rm T_a^*$ in order to
calculate the radiation 
temperature of the source $ \rm T_{mb}$. This coupling constant is 
dependent on
the spatial extent of the source. 
Taking into account the observations of standard sources
discussed above and using the suggested coupling factor of Matthews (1999),
$\eta_{\rm B}$, for a source filling the beam, we used the following
conversions from antenna
temperature to main beam temperature
for receiver A2: $\rm T_{mb}  =  T_a^* / 0.65 (\pm 5\%)$;
and for receiver B3i: $\rm T_{mb}  =  T_a^* / 0.4 (\pm 10\%)$. 
These conversion factors are slightly lower than 
the canonical values, which may have been due to minor receiver
problems at the time of the observations.

\subsection{L1689B C$^{18}$O Data}

L1689B data cubes in the
C$ \rm ^{18}$O (J=2$ \rm \rightarrow$1) and
C$ \rm ^{18}$O (J=3$ \rm \rightarrow$2) transitions
are presented in Figures 1 and 2.
The source is clearly detected in both transitions at
virtually all positions in the maps. The brightness distribution
of the core appears relatively uniform across the area mapped,
and it is not strongly peaked. The noise level per channel in the spectra
is 0.6 K for C$ \rm ^{18}$O (J=2$ \rm \rightarrow$1) and
0.4 for C$ \rm ^{18}$O (J=3$ \rm \rightarrow$2). For most positions
$ \rm T_{mb}$(peak) for C$ \rm ^{18}$O(J=2$ \rm \rightarrow$1)
lies between 5 and 6~K. The C$ \rm ^{18}$O(J=3$ \rm \rightarrow$2) 
$ \rm T_{mb}$ intensity
at most positions lies in the range 3.8 to 4.8~K. 
The most clearly detected variation in
intensity 
seems to be a drop off towards the North-East of both maps, most 
noticeably in
C$ \rm ^{18}$O(J=3$ \rm \rightarrow$2).

The full-width at half maximum (FWHM) width of the lines is 0.6--0.8km/s. The 
C$ \rm ^{18}$O(J=3$ \rm \rightarrow$2) map had too little signal to noise to
reveal any kinematic information, but a channel map of the 
C$ \rm ^{18}$O(J=2$ \rm \rightarrow$1) map was produced and is shown in 
Figure 3. The map shows no evidence of rotation which would be revealed by the
red and blue shifted maps peaking at opposite positions either side of the
centre of the core. However some evidence exists for the points 20 arcseconds
east and west of centre being blue shifted with respect to the central
position. Detailed inspection of the individual lines shows this to be true,
as they are shifted by approximately 
0.2--0.3 km/s. However, there is no clear 
evidence for line asymmetry. We 
hypothesise that this may be due to slow contraction of the core, but this
would require further observations to confirm. Using the equation for the
virial mass, $M_{\rm vir}$, of a spherical cloud (e.g. MacLaren,
Richardson and Wolfendale 1988):

\noindent
\begin{equation}
 M_{vir}=\frac{kdv^2R}{\rm G},
\end{equation}

\noindent
where v is the line width in km/s, R is the radius in pc and
k is a constant (between 126 and 210, depending on the exact form of
the density distribution),
we derive a virial mass for the core of 
$\leq$6--8$\rm M_\odot$. In Paper II we estimated the
mass within 120~arcsec to be $\geq$2.1$\rm M_\odot$, assuming
a temperature of 18K. This lower limit may be an underestimate 
since in Paper II we
used a temperature of 18K, while more recent studies (e.g. Bacmann et
al. 2000) found a temperature of $\sim$12K. This would give a mass of
$\sim$4.2$\rm M_\odot$, closer to our virial estimate. Hence we see
that the core is close to virial equilibrium.

To  quantify the spatial structure in the maps
we chose to azimuthally average the spectra in the
C$ \rm ^{18}$O (J=3$ \rm \rightarrow$2) and C$ \rm ^{18}$O
(J=2$ \rm \rightarrow$1) maps, centering the averaging around the 
C$ \rm ^{18}$O (J=2$ \rm \rightarrow$1) peak, which is coincident
with the mm-continuum peak.
In terms of angular displacement from this position, the maps contained
in each transition: 2 spectra at 7 arcsec;
4 at 16 arcsec; and 2 each at 22, 26 and 32 arcsec displacement
from the centre.

We can summarize the
results of the azimuthal averaging by saying that:
the C$ \rm ^{18}$O(J=2$ \rm \rightarrow$1) 
$ \rm T_{mb}$(peak)  at the centre is 5.5K;
the ratio of C$ \rm ^{18}$O(J=2$ \rm \rightarrow$1) at
a distance of 32~arcsec from the centre, to
C$ \rm ^{18}$O(J=2$ \rm \rightarrow$1) at centre is 1$\pm$0.1
(much less centrally peaked than the 1.3mm continuum emission); 
the ratio of C$ \rm ^{18}$O(J=3$ \rm \rightarrow$2) to
C$ \rm ^{18}$O(J=2$ \rm \rightarrow$1) is 0.78 $ \rm \pm$ 0.09
at centre; the line width is 0.7km/s $\pm$ 0.1. 
These four observations alone place strong constraints on the 
physical properties of L1689B, as we show in the next section.

In particular we wish to explore two different possibilities.
When low contrast is observed in molecular maps, and the
brightness temperature of different transitions is similar -- as is 
true in this case -- it is
often assumed that this implies that the lines are optically thick.
However another possibility is that the optical depth of each line
is constant across the map, a situation which can arise if: (i) temperature
gradients
affect the fraction of molecules in each level so as to maintain a relatively
constant column density of molecules in upper excited levels across the core;
or (ii) the molecular abundance changes so as to keep the column
averaged abundance of CO itself constant. The ratio of the line
intensities may also appear equal in these situations -- if T$_{ex}
\sim$9K the optical depth of the 
(1$\rightarrow$0) line is equal to the optical depth
of the (2$\rightarrow$1) line, 
and if T$_{ex} \sim$23K the optical depth of the (2$\rightarrow$1)
line is similar to that of the (3$\rightarrow$2) line. For the 
remainder of this
paper we use a parameterised model of the core, in combination with a
radiative transfer code to investigate these effects and others in detail,
and we
argue that a molecular abundance drop is the the most plausible explanation
for the appearance of L1689B. This abundance drop is most likely to be due
to the freeze-out of molecules onto the dust grains, giving
further evidence of the prestellar nature of L1689B.

\section{Model description}

\subsection{Model geometry of L1689B}

We have chosen a model which 
allows certain key physical characteristics
to be represented in terms of simple
analytical functions, and which at the same time incorporates
qualitatively
various physical models suggested, whilst limiting the number of free
parameters.
In this way we can parameterise
the observed appearance of L1689B and discover how the observations
constrain these parameters. We can then
hope to discover which (if any)
of the theoretical models of protostar formation 
best describes the observations of L1689B.

\begin{figure}
\begin{center}
\leavevmode
\epsfxsize=7cm
\epsfbox[-10 210 400 630]{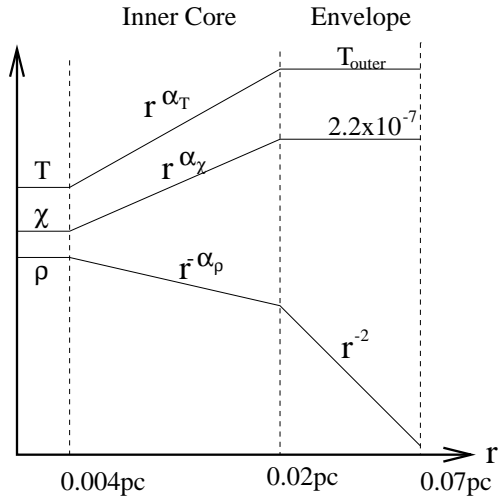}
\end{center}
\caption{The model core for L1689B, partly based on the findings of
Paper II. The core has an outer envelope with a density profile
which falls as $ \rm r^{-2}$. The outer envelope is isothermal and is 
assumed to have the CO abundance typical of the Ophiuchus cloud. The 
inner core region has a more flattened density profile as suggested by the
mm/submm continuum observations. The inner core is not assumed to be 
isothermal, and the CO abundance can also vary. In order to
keep all parameters finite, the core has an inner radius of 0.004pc, well 
below any current single dish telescope resolution, within which
all physical conditions are homogeneous.}
\end{figure}

The model parameters are shown schematically in Figure 4.
We first make the simplification that L1689B is spherically symmetric, 
which significantly simplifies the radiative transfer
analysis.
Given that the ellipticity of L1689B appears to be low (Paper II), we
feel this justifies the considerable saving of coding and cpu time.
The model core
consists of an isothermal envelope of outer radius 0.07pc  
surrounding
an inner core of radius 0.02pc (c.f. Paper II). The
temperature and CO abundance are kept constant in the envelope, and
the density varies as r$^{-2}$. Within the inner core, all three
physical parameters of density, temperature and abundance are allowed
to vary according to a power-law dependence, as shown in Figure 4.
The radius at which a break in the power law density profile is observed
(the break between the inner core and outer envelope)
will also be referred to as the critical radius (c.f. R$_{flat}$ in Paper II).
To ensure that the parameter values remain
finite we also set an inner radius of the core equal to 0.004pc (well within
the resolution of JCMT or IRAM), which does not affect the results.

The inner core density profile, 
temperature profile and CO abundance profile 
power-laws are given by
$ \rm r^{-\alpha_{\rho}}$,
$ \rm r^{\alpha_{T}}$,
and $ \rm r^{\alpha_{\chi}}$ respectively. 
Hence there are 5 free parameters in the model: 
the central density; the outer temperature; and the value of the three
power-law indices. The critical radius between core and envelope is
set by the mm continuum data (Paper II). The outer abundance is set to the
canonical C$^{18}$O abundance for 
Ophiuchus.

\subsection{The $\lambda$-Iteration Code}

We used a previously tried and tested radiative transfer code,
known as the Stenholm Code (Stenholm 1977, Matthews 1986), which uses a 
$\lambda$-iteration method to
solve for the population levels
and produce model spectra output.
The code models a spherically symmetric molecular cloud as a set of shells,
each with uniform physical conditions. For each shell it requires that the
$ \rm H_2$ density, the $ \rm H_2$ temperature and the abundance of 
C$ \rm ^{18}$O
(with respect to $ \rm H_2$) be specified. 
Because molecular line widths are generally greater than
predicted from simple thermal line broadening, a micro-turbulent component
of velocity, $\delta v$, can also be specified for each shell.

\begin{figure*}
\centering
\leavevmode
\epsfxsize=12cm
\epsfbox[0 0 420 285]{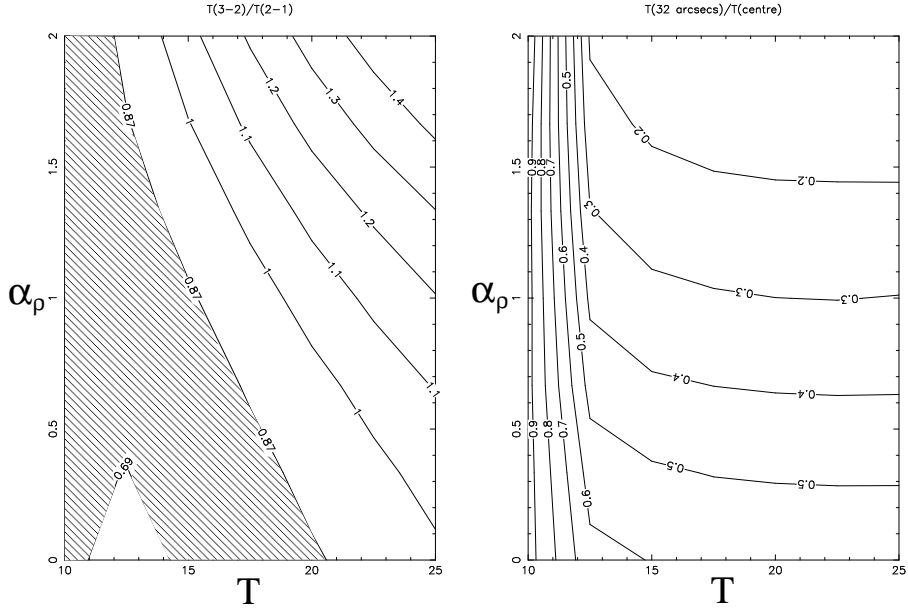}
\caption{Two contour plots illustrating the effects of varying
$\alpha_\rho$ and T, whilst setting $\alpha_T =\alpha_\chi = 0$
(i.e. isothermal models). (a) Contours of the values
of the ratio of $\rm T_{mb}$(J=3$ \rm \rightarrow$2) to 
$\rm T_{mb}$(J=2$ \rm \rightarrow$1). The area of this plot that is
consistent with the data shown in Figures 1 \& 2 is shaded. 
(b) Contours of the values of the ratio of 
$\rm T_{mb}$ at a radius of 32 arcsec from the centre to
$\rm T_{mb}$ at core centre for the C$^{18}$O (J=2$\rightarrow$1)
transition. The value observed in the L1689B data
for this ratio was 1.0$\pm$0.1.
No physically realistic value of $\alpha_\rho$ can reproduce this
value of the ratio.}
\end{figure*}

The code starts by setting the population levels of the C$ \rm ^{18}$O
rotationally excited states to be consistent with the Boltzmann distribution.
The equation of radiative transfer is then numerically 
solved to obtain  $ \rm I_{\nu}$, the radiation field in each shell. 
Using this
radiation field the rate equation is 
then solved to give a new, modified set of population levels. 
The procedure of calculating $ \rm n_J$ (the population levels of the
C$ \rm ^{18}$O) followed by $ \rm I_\nu$ (the radiation field)
is repeated until a stable solution is approached and convergence
is achieved.
The code can then solve 
the radiative transfer equation for lines of sight through the cloud and 
simulate observed line profiles by smoothing with a Gaussian spatial profile
similar to the beam profile of the JCMT.
The $ \rm H_2$ collision rates for the model were obtained from Flower \& 
Launey (1985), who calculated the collision rates for para-$H_2$ up to the 
$ \rm J=11$ rotational level, for temperatures between 10
and 250K, and for ortho-$H_2$ up to the $ \rm J=6$ level between 10 and
100K. 

Throughout the simulations (unless stated otherwise) the model core
representing L1689B had the following properties:
The model core consisted of 20 logarithmically spaced shells (giving  
increased resolution towards the centre of the core),
and a critical radius $ \rm r_{crit}=0.02pc$ marking the boundary
between the core and envelope.
The micro-turbulent velocity component for each shell throughout the cloud
was given by $ \rm \delta v(r) \propto r^{0.3}$, up to a maximum
0.8~kms$^{-1}$ at the 
outside. This was 
chosen to be consistent with the observations summarised
by Larson (1981) and to fit the 
observed line widths. 
The ortho- to para-$H_2$ ratio was fixed at 1.

The following section investigates the effect of varying the free 
parameters in the model of L1689B:
$ \rm \alpha_{T}$; $ \rm \alpha_{\rho}$; $ \rm \alpha_{\chi}$; and
$ \rm T_{outer}$.
The central density, $ \rm \rho_c$, 
was normalized in the simulations so as to 
reproduce the central brightness of the core in the 
(2$\rightarrow$1) transition.

\section{Model Results}

\subsection{Isothermal Models}

The simplest models of cloud cores assume a constant temperature. So we
first investigate how the predicted appearance of L1689B is dependent 
on the inner core radial
density profile $ \rm \rho \propto r^{-\alpha_{\rho}}$ if a single
temperature, T (=T$_{outer}$), 
exists throughout the core. In other words, we set
both $\alpha_T$ and $\alpha_\chi$ to zero. Hence,
the predicted appearance for the set of cores with 
temperatures between 10 and
25K and $ \rm \alpha_{\rho}$ between 0 and 2 was calculated. 
By the nature of this form of modelling,
we can run whole families of models and produce large output data-sets.
We then illustrate these models as a series of contour plots in the
two-dimensional phase space of the two parameters we are varying.

Figure 5 shows two contour plots that illustrate the effects of varying
$\alpha_\rho$ and T. Figure 5(a) shows contours of the values
of the ratio of $\rm T_{mb}$(J=3$ \rm \rightarrow$2) to 
$\rm T_{mb}$(J=2$ \rm \rightarrow$1). The area of this plot that is
consistent with the data shown in Figures 1 \& 2 is shaded. 
Figure 5(b) shows contours of the values of the ratio of 
$\rm T_{mb}$ at a radius of 32 arcsec from the centre to
$\rm T_{mb}$ at core centre for the C$^{18}$O (J=2$\rightarrow$1)
transition. As noted in section 2 above, the value observed in the data
of L1689B for this ratio was 1.0$\pm$0.1.
It can be seen from Figure 5(b) that for T$>$13K this value increases as
$\alpha_\rho$ decreases and that the brightness profile in this area is
mainly dependent on the density profile of the core. However for T$<$13K
the brightness profile becomes strongly dependent on T. This is because
the C$^{18}$O transitions become increasingly optically thick.

The reason why the
(J=3$\rightarrow$2) to (J=2$\rightarrow$1)
brightness ratio of the model core 
increases with both T and $\rm \alpha_\rho$
is because the upper rotational levels of C$ \rm ^{18}$O
become more populated as the temperature and density increase. 
It should be noted that the singular isothermal sphere model
represents the upper border of Figure 5(b). This is the part of
the plot furthest from consistency with the data for T$>$13K.
This agrees with the conclusions of Papers I \& II that L1689B
does not represent the singular isothermal sphere initial conditions 
for star formation.

Although at first sight setting T=10K seems to offer a possible explanation
for L1689B's brightness profile, another effect led us to reject this as a
possible model for L1689B. At 10K the model has very high optical depths and
the predicted line profiles are significantly wider and show an extended 
`flat top'. This is illustrated in Figure 6. The central 
C$ \rm ^{18}$O line profile is shown as observed (the data points) and as
predicted by the model $\alpha_\rho$=0, T=12K (solid line) and $\alpha_\rho$=0,
T=10K (dashed line). It is clear that the T=10K model does not agree with the
data. We therefore need to find some other mechanism to explain the low
contrast observed in the data.

\begin{figure}
\begin{center}
\leavevmode
\epsfxsize=6cm
\epsfbox[60 20 370 370]{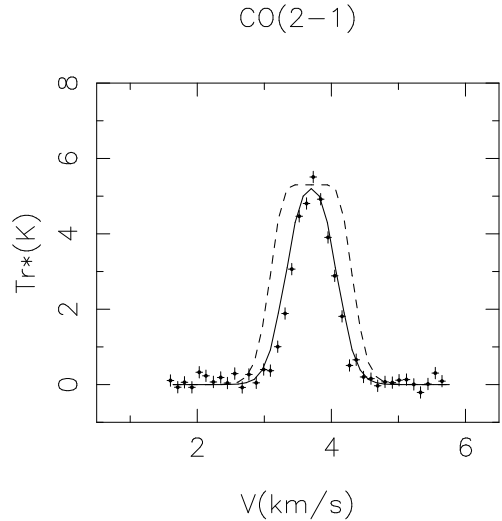}
\caption{Comparison of observed C$ \rm ^{18}$O (J=2$ \rm 
\rightarrow$1) line (data points) with two isothermal models. The first
model (solid line) has T=12K, the second (dashed) has T=10K. The second
clearly shows a flat top due to high optical depths, and is not consistent
with the data.}
\end{center}
\end{figure}

\subsection{Varying temperature models}

\begin{figure*}
\centering
\leavevmode
\epsfxsize=12cm
\epsfbox[0 400 550 780]{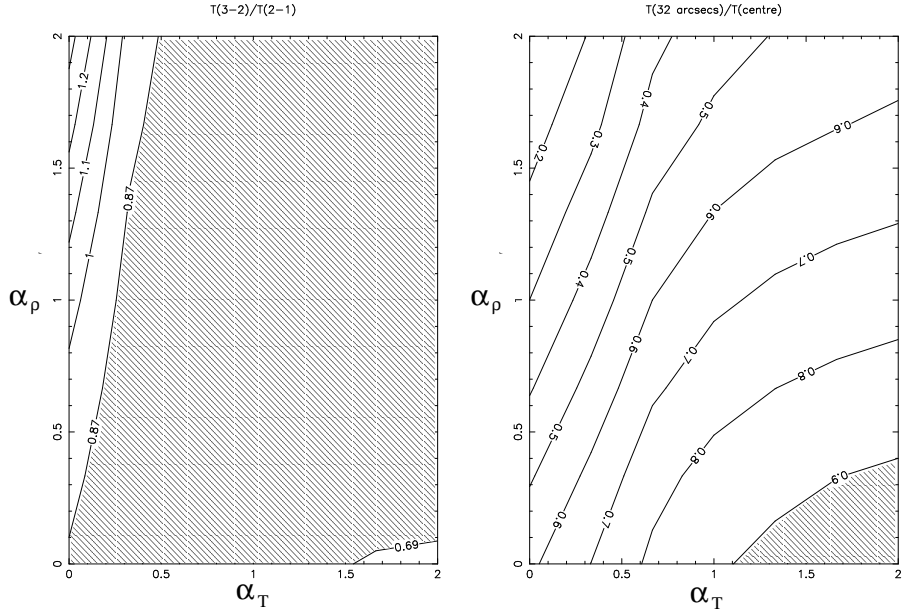}
\caption{Two contour plots illustrating the effects of varying
$\alpha_\rho$ and 
$\alpha_T$ simultaneously. (a) Contours of the values
of the ratio of $\rm T_{mb}$(J=3$ \rm \rightarrow$2) to 
$\rm T_{mb}$(J=2$ \rm \rightarrow$1). The
shaded area is consistent with the L1689B data. 
(b) Contours of the values of the ratio of 
$\rm T_{mb}$ at a radius of 32 arcsec from the centre to
$\rm T_{mb}$ at core centre for the C$^{18}$O (J=2$\rightarrow$1)
transition. The shaded area is consistent with
the data from Figures 1 \& 2. 
Note that the two shaded areas overlap slightly.}
\end{figure*}

We next
investigated a set of models in which the outer envelope was at a
temperature of 20K, but in the inner region 
the temperature profile falls as 
$ \rm T \propto r^{\alpha_{T}}$ to a minimum temperature
of 10K. This means that, when $ \rm \alpha_T$ is
greater than approximately 0.5, an inner 10K region is created, whose size
increases as  $ \rm \alpha_T$ rises.  When $ \rm \alpha_T$ is
2.0, the radius of this region is 0.015pc.  

Figure 7 shows two contour plots that illustrate the effects of varying
$\alpha_\rho$ and 
$\alpha_T$ simultaneously. Figure 7(a) shows contours of the values
of the ratio of $\rm T_{mb}$(J=3$ \rm \rightarrow$2) to 
$\rm T_{mb}$ (J=2$ \rm \rightarrow$1). The area of this plot that is
consistent with the L1689B data is shaded. 

Figure 7(b) shows contours of the values of the ratio of 
$\rm T_{mb}$ at a radius of 32 arcsec from the centre to
$\rm T_{mb}$ at core centre for the C$^{18}$O (J=2$\rightarrow$1)
transition. As noted in section 2 above, the value observed in the data
of L1689B for this ratio was 1.0$\pm$0.1.
Once again the part of this plot that is consistent with
the data from Figures 1 \& 2 is shown as a shaded area on Figure 7(b).
The fact that figure 7(b) shows a much larger variation than Figure 7(a)
illustrates the importance of taking observations at more than one radius
if one wishes to truly characterize cloud cores, and shows clearly that
single spectra at the centre of a cloud do not constrain its parameters.

This time there is a set of values of $\alpha_\rho$ and 
$\alpha_T$ that is consistent with all of the C$^{18}$O data.
This approximately corresponds to $1.1 < \alpha_T < 2$
and  $0 < \alpha_\rho < 0.4$. However, at the 1$\sigma$ level,
we can also say that if $\alpha_\rho\leq 0.1$, then $\alpha_T$
must by $\leq1.5$.
These results reveal that a decrease in temperature
towards the centre of the core does indeed help
to explain the C$ \rm ^{18}$O appearance of L1689B.
In fact a model which has a flat inner density
profile and a very steep temperature drop from 20K at a radius of 0.02pc,
to 10K at a radius of 0.015pc, is consistent with our observations.

\begin{figure}
\begin{center}
\leavevmode
\epsfxsize=4cm
\epsfbox[0 0 180 350]{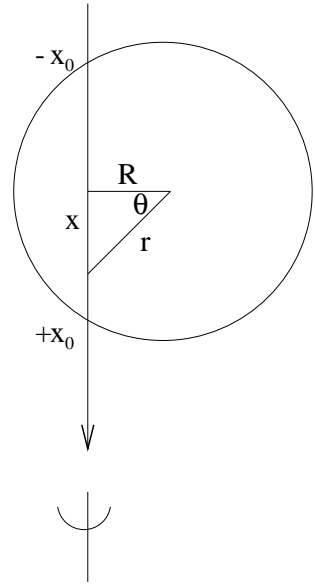}
\caption{Sketch of the assumed observing geometry of L1689B.}
\end{center}
\end{figure}

\subsection{Continuum data revisited}

A 1.3-mm continuum map of L1689B was published in Paper II. In this section
we incorporate those observations to see what further constraints we can
place on our model of L1689B. Remembering that the
millimetre continuum arises from thermal emission from
dust grains, the flux density in this situation
(e.g. Hildebrand 1983) is given by:

\begin{equation}
 S_{\nu}= B_{\nu}(T) \Omega (1 - e^{-\tau_\nu}),
\end{equation}

\noindent
where $B_{\nu}(T)$ is the black-body function at temperature $T$, 
$\Omega$ is the source solid angle and $\tau_\nu$ is the optical depth.
This form is often referred to as a grey-body function.
The black-body function can often be simplified to the so-called
Rayleigh-Jeans approximation in cases where $(h\nu/kT)<<1$. However,
at 1.3 mm, $(h\nu/k)=10.8K$, 
which is within a factor of 2 of our fitted
temperature for the the L1689B core (see above). Hence, in this case 
the Rayleigh-Jeans approximation may introduce errors, so we here
outline a somewhat more rigorous approach.

Figure 8 shows our assumed observing geometry.
We still make the assumption that the dust properties, such as
grain size distribution, abundance with respect to H$_2$ and emissivity,
are constant throughout the cloud. Then we see that 
the intensity arising from an element of the cloud of
length $dx$ at a given frequency is:

\begin{equation}
dI(x) \propto \rho(x)B(T(x))dx,
\end{equation}

\noindent
where $\rho(x)$ is the density profile along the line of sight 
and $T(x)$ is 
the temperature profile of the dust along the line of sight --
see Figure 8.

\begin{figure}
\setlength{\unitlength}{1mm}
\begin{picture}(60,80)
\includegraphics{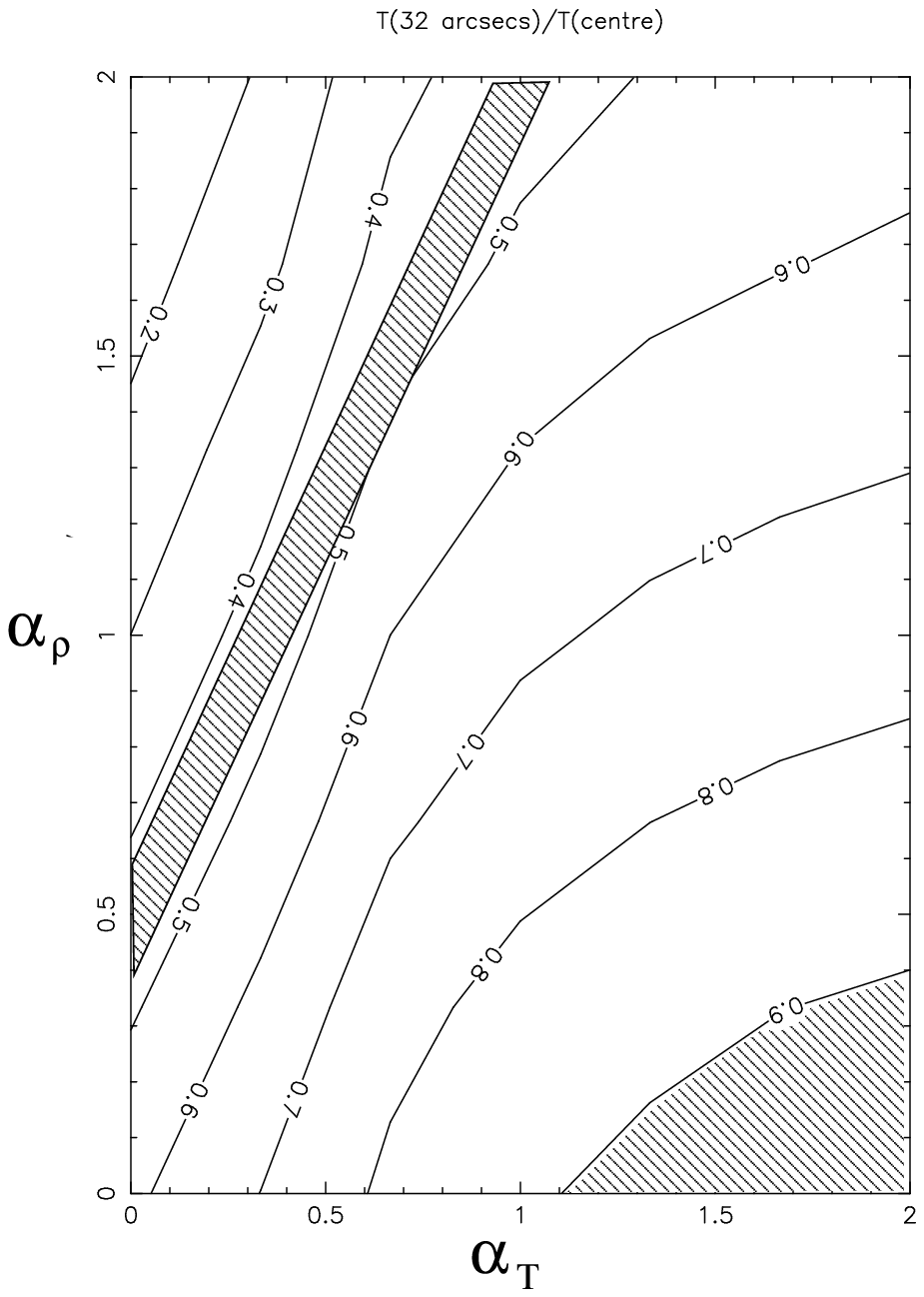}
\end{picture}
\caption{Contours of the values of the ratio of 
$\rm T_{mb}$ at a radius of 32 arcsec from the centre to
$\rm T_{mb}$ at core centre for the C$^{18}$O (J=2$\rightarrow$1)
transition. The shaded area at lower right
is consistent with the data from Figures 1 \& 2.
The shaded strip towards the upper left indicates the constraints
placed on $\alpha_T$ and $\alpha_\rho$ by the continuum data.
Note that the two shaded areas do not overlap.}
\end{figure}

The simple form often used for power-law radial profiles, is
that if $T(r) \propto r^{-q}$, $\rho(r) \propto r^{-p}$, 
and the flux density $S_{\nu}(\theta) \propto \theta^{-m}$
then the indices are related by the simple expression $m=p+q-1$ (Casali 1986,
Adams 1991).
This is derived by assuming the depth of the cloud remains constant across the
entire core (e.g. a slab or infinite cloud). 
However, this form only holds for the Rayleigh-Jeans approximation, 
and for the case where the radial density and temperature each follows
a single power-law, and there is no break in the power laws, such as
we have in our model -- see Figure 4. We derive below a more rigorous
set of relations.

If the density profile 
$\rho(r)$ and the temperature profile $T(r)$
are known, 
where r is the radial distance from the cloud centre, then the observed
surface brightness I(R)
is given by the equation of radiative transfer in the
assumption of optically thin emission:

\begin{equation}
I(R) \propto \int_{-x_0}^{+x_0}\rho(r)B(T[r])dx.
\end{equation}

\noindent
Making the substitutions 
$R=r\cos \theta$ and $dx=[R d\theta / \cos^2 \theta]$
we derive:

\begin{equation}
I(R) \propto R \int_{-\cos^{-1} (R/R_c)}^{\cos^{-1} (R/R_c)} 
\rho(R/ \cos \theta) \
B(T[R/ \cos \theta]) \frac{d \theta}{\cos^2 \theta}.
\end{equation}

\noindent
For the model of L1689B, the density profile in the outer envelope
is represented by the power law 
$\rho \propto r^{-2}$. Hence, for $R>r_{crit}$ 
(i.e. for observations of the outer envelope) we have:

\begin{equation}
I(R) \propto R^{-1} \int_{-\pi/2}^{\pi/2} 
(\cos \theta )^{-4} d\theta,
\end{equation}

\noindent
where the limits in the integration have been set
using the approximation that the core extends to infinity.
This integral is independent of R and therefore
we predict the $I(R) \propto R^{-1}$
behaviour in the envelope of L1689B 
that was actually observed in the continuum in Paper II.

However, for the inner 
core the equation for the projected surface 
brightness becomes more complicated:

\begin{eqnarray}
\lefteqn{I(R) \propto \frac{B(T_{env})}{\omega} 
\left( \frac{\pi}{2}\cos^{-1}(\omega)
\right) +} \\ \nonumber
 & \omega\int_0^{\cos^{-1}(\omega)}\frac{\omega^{-\alpha_\rho}
B\left(T_{env}\omega^{-\alpha_T}
cos^{\alpha_{T}}\theta\right)cos^{\alpha_{\rho}}\theta}{cos^2\theta}
d\theta,
\end{eqnarray}

\noindent
where we have used the substitution
$\omega=R/r_{crit}$, and $\rm T_{env}$ is the envelope temperature. The first
term of this equation is the contribution to the signal from the envelope,
while the second term is the contribution from the core.
This equation can be solved numerically for
any value of $\alpha_{\rho}$ or $\alpha_{T}$, to predict the
brightness profile I(R). 

Numerical solution of equation 7 in the isothermal case
allows us to compare the continuum radial flux density variation observed in
Paper II, with that predicted by the parameterised model we have presented
here. We find that models with 
$\rho\propto r^{-0.5\pm0.1}$ (i.e. $\alpha_{\rho}=0.5\pm0.1$) best match
the data presented in Paper II.
A more general, non-isothermal,
solution to the continuum appearance of L1689B can be expressed
approximately as:

\begin{equation}
\alpha_{\rho} - 1.5\alpha_{T} = 0.5 \pm 0.1.
\end{equation}

\noindent
This solution was found by numerically solving equation 7 for all values
of $\alpha_{\rho}$ between 0 and 2, and $\alpha_{T}$ between 0 and 2,
and then finding which range of values give solutions best matching the data
in Paper II.
This is a new and different constraint to that discovered in the above
modelling of the C$^{18}$O data.

\begin{figure*}
\centering
\leavevmode
\epsfxsize=12cm
\epsfbox[0 0 420 280]{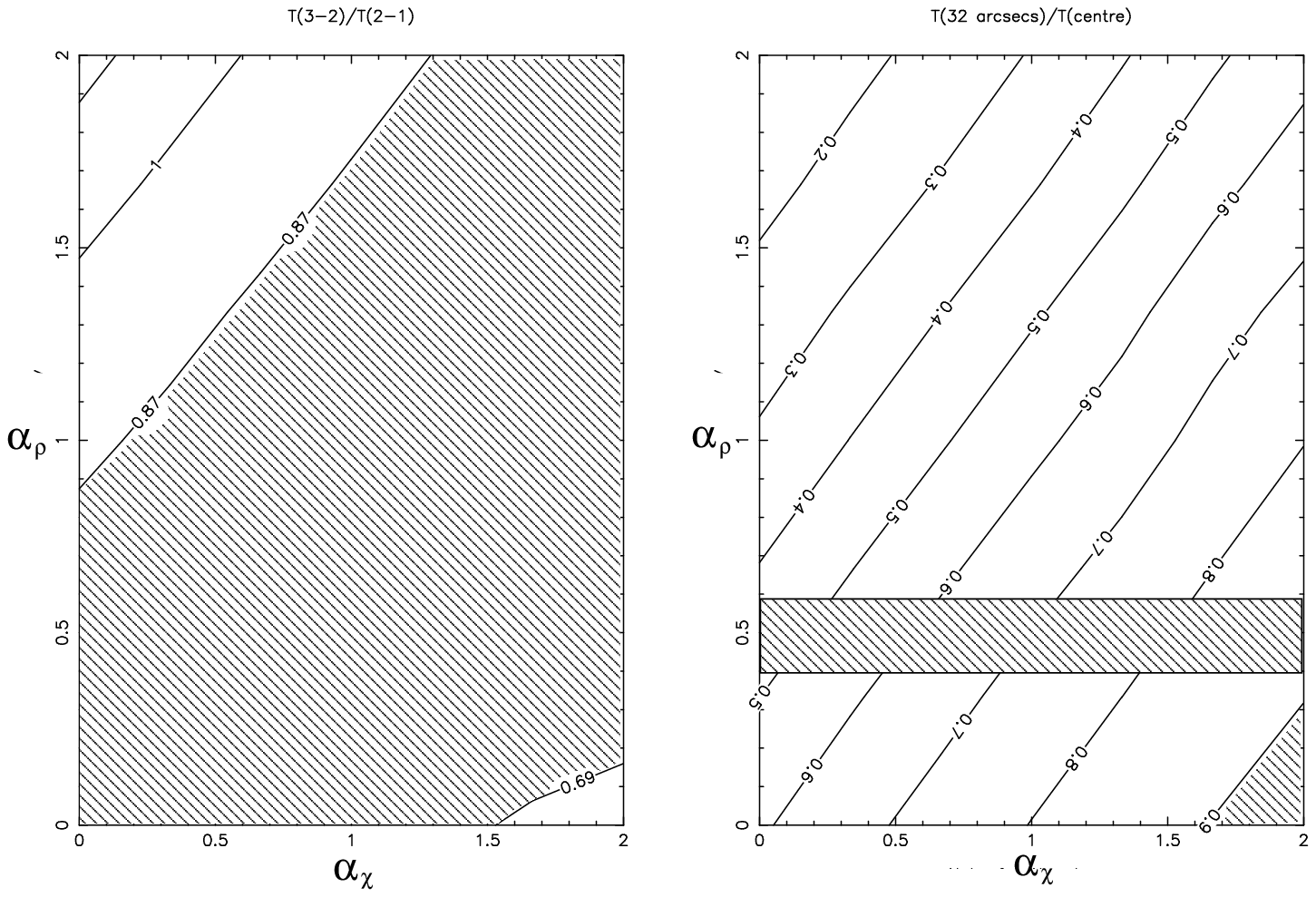}
\caption{Two contour plots illustrating the effects of varying
$\alpha_\rho$ and 
$\alpha_\chi$ simultaneously. (a) Contours of the values
of the ratio of $\rm T_{mb}$(J=3$ \rm \rightarrow$2) to 
$\rm T_{mb}$(J=2$ \rm \rightarrow$1). The
shaded area is consistent with the L1689B CO data. 
(b) Contours of the values of the ratio of 
$\rm T_{mb}$ at a radius of 32 arcsec from the centre to
$\rm T_{mb}$ at core centre for the C$^{18}$O (J=2$\rightarrow$1)
transition. The shaded area at the lower right hand
corner is consistent with
the data from Figures 1 \& 2. 
The horizontal shaded strip indicates the constraints
placed on $\alpha_\rho$ by the continuum data.}
\end{figure*}

Figure 9 illustrates this additional constraint introduced by the
continuum data by reproducing the contours from
Figure 9(b) -- the values of the ratio of 
$\rm T_{mb}$ at a radius of 32 arcsec from the centre, to
$\rm T_{mb}$ at core centre for the C$^{18}$O (J=2$\rightarrow$1)
transition -- as a function of varying $\alpha_{\rho}$ and $\alpha_{T}$.
Once again the part of this plot that is consistent with
the CO
data from Figures 1 \& 2 is shown as a shaded area on Figure 7(b).
However, now we have added an additional shaded strip at the
upper left of the plot illustrating the constraint placed by the
continuum data, as described by equation 8.
We see that the two shaded areas do not overlap, indicating that
there is no simultaneous fit to the CO data and the continuum data.
Hence, a more complex model is required to simultaneously fit
all of the data.
 
\subsection{Varying the CO abundance}

The simplest way to try to reconcile the CO data to the continuum
data is to vary the abundance of gas-phase CO relative to H$_2$.
This is predicted to occur in regions of high density, as the CO
freezes out onto the dust grains, and it has been observed qualitatively
in many regions on many occasions (e.g. Mezger et al. 1990; 
Gibb \& Little 1998;
Ward-Thompson et al. 2000). It is thought to occur at temperatures 
less than 17K (Nakagawa 1980, Bergin \& Langer 1997). So we ran a set of
models to investigate the effect of decreasing the CO abundance
towards the centre of the core. These models had a single temperature
of 20K, an inner density gradient as before, of 
$\rho \propto r^{-\alpha_{\rho}}$, and a
CO abundance fraction $\chi \propto r^{\alpha_{\chi}}$. We
then investigated the regime $0<\alpha_{\rho}<2$ and
$0<\alpha_{\chi}<2$.

Figure 10 shows two contour plots that illustrate the effects of varying
$\alpha_\rho$ and 
$\alpha_\chi$ simultaneously. Figure 10(a) shows contours of the values
of the ratio of $\rm T_{mb}$(J=3$ \rm \rightarrow$2) to 
$\rm T_{mb}$(J=2$ \rm \rightarrow$1). The area of this plot that is
consistent with the CO data is shaded. 

Figure 10(b) shows contours of the values of the ratio of 
$\rm T_{mb}$ at a radius of 32 arcsec from the centre to
$\rm T_{mb}$ at core centre for the C$^{18}$O (J=2$\rightarrow$1)
transition. As noted in section 2 above, the value observed in the data
of L1689B for this ratio was 1.0$\pm$0.1.
Once again the part of this plot that is consistent with
the data from Figures 1 \& 2 is shown as a shaded area in
the lower right hand corner of Figure 10(b).

Including the continuum data in this case, with $\alpha_T=0$,
equation 10 simplifies to $\alpha_{\rho} = 0.5 \pm 0.1$.
Figure 10(b) illustrates this additional constraint as
an horizontal shaded strip in the lower half of the plot. Again
we see that the two shaded areas do not overlap, indicating that
there is no simultaneous
formal fit to the CO data and the continuum data.
However, this is only at the 1$\sigma$ level, and we find that the
shaded regions would overlap at the 2$\sigma$ level for the values
$\alpha_\chi \sim 1.9-2$ and $\alpha_\rho \sim 0.4$.
This abundance effect is quite extreme -- it implies
a 95\% drop in C$^{18}$O abundance from edge to centre --
although not unprecedented. Gibb and Little (1998) found a similar reduction
in the C$^{18}$O abundance in HH23--26.
 
Furthermore, if there is such a large amount of freeze-out taking place,
this will increase the average grain size towards the core centre and
consequently the dust grain mass opacity. In this case we would 
over-estimate the central density profile exponent, $\alpha_\rho$
derived from the continuum data. This would then act to shift the
horizontal shaded region in Figure 10(b) downwards, causing a genuine
overlap region at the 1-$\sigma$ level. Thus we have a fit to the
data with $\alpha_\chi\sim 2$ and $\alpha_\rho\leq 0.4$. The spectra
produced by the model were compared with those observed 
and there was good agreement within the errors. This is 
an isothermal fit. However we cannot rule out a slight temperature
gradient, depending upon how much the dust opacity parameters that are
not well constrained can vary.

We ran this set of simulations for several temperatures 
from 10 to 25K. We found that the
predicted surface brightness did not vary significantly with T, even 
at very low temperatures. However,
the predicted ratio of C$^{18}$O (J=3$\rightarrow$2) to C$^{18}$O
(J=2$\rightarrow$1) did fall with T, and for models 
with T$<$14K the predicted
values were too low to match our observations.

\subsection{Other factors}

There are other factors than those discussed above that may affect
the conclusions of the modelling. One such factor is that
the ortho- to para-H$_2$ ratio in 
these cores is not well defined. Since
ortho-H$_2$ has a higher cross section for 
collisions with CO, the ratio of ortho- to para-H$_2$ may
affect the conclusions drawn. 
It was found that by increasing the fraction of ortho-H$_2$, 
the C$^{18}$O centre-to-edge ratio increased. However, it hardly
affected the ratio of the
the (J=3$\rightarrow$2) to (J=2$\rightarrow$1) transitions.
Varying the ortho- to para- ratio altered the centre-to-edge
ratio by 0.1 at most.

Similarly, the precise form of the microturbulence profile may influence 
the appearance of the core. For example, it was found that when
using a constant $\delta v$ microturbulence profile, 
the brightness gradient decreased slightly --
T(32 arcsec)/T(centre) increased by $\sim$ 0.1.
However, neither of these two factors affected 
the appearance of the L1689B data strongly
enough to affect the conclusions drawn above.

The geometry of the core as a whole may also change the exact form of the
solutions. In particular, any departure
from spherical symmetry would have an effect.
Likewise, any variations from
homogeneity -- i.e. clumpiness within the core -- is also likely to
affect the predicted appearance of the core in mm/submm observations. 
It is difficult to quantify these effects, and further data
would be needed to constrain these extra parameters.

We also investigated the possible effect of having a small temperature gradient
in the envelope, or allowing the density gradient in this region to fall less
steeply ($\rho \propto n^{-1.7}$). This was found to only vary the ratio of 
T(32 arcsec)/T(centre) by $\sim$ 0.05. Checks to ensure that the
exact values of the inner radius, and the cloud size did not significantly
affect our results were also made.

\section{Conclusions}

In this paper we have presented new C$^{18}$O (J=3$\rightarrow$2) and
(J=2$\rightarrow$1) data of the pre-stellar core L1689B. We have used
a spherically symmetric radiative transfer code to model these data in
terms of three parameters -- the gradients in temperature, density and
C$^{18}$O abundance.
In the three dimensional parameter space defined by the three exponents,
$\alpha_\rho$, $\alpha_T$, and $\alpha_\chi$, we have found 
solutions consistent with the data.
There is a region
defined by $\alpha_\rho - 1.5 \alpha_T=0.5\pm0.1$, which is consistent 
with the mm continuum observations of L1689B.

There is also a set of solutions that can simultaneously predict the 
molecular CO appearance and the mm continuum data, implying
$\alpha_\rho\leq 0.4, \alpha_T=0$ (although this is not well constrained)
and $\alpha_\chi =2$. Hence, we have shown that there is freeze-out of
CO towards the centre, and
we have constrained the radial density and temperature profiles.
Thus we have shown that
the combination of mm/submm spectral line and continuum data, with a
rigorous treatment of radiative transfer, is a powerful method of
investigating the physics of pre-stellar evolution.

\section*{Acknowledgments}

N.E.J. wishes to acknowledge PPARC for studentship support while this
research was carried out. The authors
would also like to thank the JCMT telescope operators for their
assistance during these observations, and Les Little for
providing the original version of the model code that was modified
for use in this work. We also thank
the referee Andy Gibb, who made several helpful comments and
suggestions.

\label{lastpage}

\end{document}